\documentclass[showpacs,preprintnumbers,amsmath,amssymb,twocolumn]{revtex4}

\usepackage{epsfig}

	     \usepackage{graphicx}

\usepackage{dcolumn}

\usepackage{bm}
\newcommand{\be}{\begin{equation}}

\newcommand{\ee}{\end{equation}}\newcommand{\ba}{\begin{eqnarray}}

\newcommand{\ea}{\end{eqnarray}}

\begin{document}

\title{Four Quark Interpretation of $Y(4260)$}

\author{L. Maiani}\email{luciano.maiani@roma1.infn.it}
\author{V. Riquer}\email{veronica.riquer@cern.ch}
\affiliation{Universit\`{a} di Roma `La Sapienza' and I.N.F.N., Roma, Italy}

\author{F. Piccinini}\email{fulvio.piccinini@pv.infn.it}
\affiliation{I.N.F.N. Sezione di Pavia and Dipartimento di Fisica Nucleare 
e Teorica, via A.~Bassi, 6, I-27100, Pavia, Italy}

\author{A.D. Polosa}\email{antonio.polosa@cern.ch}
\affiliation{Dip. di Fisica, Universit\`{a} di Bari and I.N.F.N., Bari, Italy}

\date{\today}

\begin{abstract}
We propose that the $Y(4260)$ particle recently announced by BaBar
is the first orbital excitation of a diquark-antidiquark state 
$\left([cs][\bar{c}\bar{s}]\right)$. 
Using parameters recently determined to describe the $X(3872)$ and $X(3940)$ 
we show that the $Y$ mass is compatible with the orbital excitation 
picture. A crucial prediction is that $Y(4260)$
should decay predominantly in $D_s\bar{D}_s$. The $Y(4260)$ should 
also be seen in $B$ non-leptonic decays in association with one kaon.
We consider the full nonet 
of related four-quark states and their predicted properties. Finally,
we comment on a possible narrow resonance in the same channel.

ROMA1-1408/2005, FNT/T-2005/07, BA-TH/516/05

\end{abstract}

\maketitle


In a series of exciting experiments, BELLE and BaBar have discovered several 
states that, although decaying in charmonium plus pions, do not seem to fit 
the $c\bar c$ picture, in particular the $X(3872)$ and $X(3940)$ states.

In a recent paper~\cite{xparticles} we have pointed out that the properties 
of the new states  can be well explained if they are 
S-wave diquark-antidiquark bound states with the composition ($q=u,d$):
$[(cq)(\bar c\bar q)]_{{\rm S-wave}}$.
An alternative scenario is the molecular picture where the $X(3872)$ would be
a $D^0D^{*0}$ bound state. A crucial difference between the two alternatives
is that colored objects
in a rising confining potential, such as diquarks,
should exhibit a series of orbital angular momentum excitations. 
This is clearly at variance with the molecular picture. Colorless 
objects bound by a short range potential should have a 
very limited spectrum, possibly restricted to  S-wave states only.

In this note we would like to propose that the first orbital excitation of a 
diquark-antidiquark state may have indeed been found in the
state $Y(4260)$ recently announced by the BaBar collaboration~\cite{babar}. 
We discuss the properties of the new state in this framework and spell out 
a few distinctive predictions. The most revealing among them 
is that the dominant decay mode of $Y(4260)$ should be in $D_s\bar{D}_s$ 
pairs. We shall also briefly discuss other states implied by the scheme
and their properties. We comment on the possibility of an additional 
narrow state.

The $Y(4260)$ is observed by BaBar in $e^+e^-$ annihilation, 
in association with an 
Initial-State-Radiation photon, which implies $J^{PC}=1^{--}$. The particle
has a width of about $90$~MeV and it is seen to decay in 
$J/\psi\;\pi^+\pi^-$. The $\pi^+\pi^-$ mass distribution peaks around $1$~GeV,
consistently with a decay
into $J/\psi\; f_0(980)$.
BaBar reports the value~\cite{babar}:
\begin{equation}
\Gamma(Y\to e^+e^-)\times Br(Y\to J/\psi\pi^+\pi^-)=5.5\pm 
1.0^{+0.8}_{-0.7}~{\rm eV}
\label{eq:babar}
\end{equation}

The diquark-antidiquark assumption together with the negative parity
call for at least one unit of orbital angular momentum. In addition, 
the decay into $f_0(980)$, which
fits the
$\left([sq][\bar s\bar q]\right)_{\rm{S-wave}}$ hypothesis~\cite{prl}, 
suggests a 
$[cs][\bar{c}\bar{s}]$ composition. All considered, we are led to 
the following assumption for the $Y(4260)$:
\begin{equation}
\label{eq:uno}
Y(4260)=\left([cs]_{S=0}[\bar{c}\bar{s}]_{S=0}\right)_{\rm{P-wave}}
\end{equation}
with both diquarks in a  $\bf{\bar{3}}$ color state. 

As discussed 
in~\cite{xparticles} we expect diquarks involving charmed quarks to be bound 
also in states 
with non-vanishing spin ({\it bad diquarks}~\cite{jaffe}, 
with $S=1$). 
Thus, several other states with $J^{PC}=1^{--}$ are possible and one would 
expect the physical $Y(4260)$ to be a linear superposition of all such states.
The state in~\eqref{eq:uno} is supposedly the lowest lying among them 
and we restrict to it in this first analysis. 

Following~\cite{xparticles}, a simple mass formula for the $Y$ state can be 
given as follows:
\begin{equation}
\label{eq:due}
M_Y=2m_{[cq]}+2(m_s-m_q)-3 \kappa_{cs}+B_c\left(\frac{L(L+1)}{2}\right).
\end{equation}
$m_{[cq]}$ is the mass of the heavy-light diquark as computed 
in Ref.~\cite{xparticles}, i.e., $m_{[cq]}=1933$~MeV, $m_q$ and $m_s$ are the 
constituent up and strange quark masses, respectively. A fit to the lowest 
lying meson and baryon masses, as reported in~\cite{xparticles}, 
gives $m_s-m_q=185$~MeV. Spin-spin interactions 
are described by the Hamiltonian:
\begin{equation}
H_{\rm{spin-spin}}=2\kappa_{cs}(\vec{S}_c\cdot\vec{S}_{s}+
\vec{S}_{\bar{c}}\cdot\vec{S}_{\bar{s}})
\end{equation}
and $-3\kappa_{cs}$ is its eigenvalue in the $S=0$ state. The value of 
$\kappa_{cs}$ is obtained from a fit to the charmed strange
baryon spectrum and is reported in~\cite{xparticles} as 
$(\kappa_{cs})_{\bf{\bar{3}}}=25$~MeV.
In Eq.~\eqref{eq:due} we are neglecting
spin-spin interactions between quarks and antiquarks
(because of the angular momentum barrier which separates the diquark
from the antidiquark) and the spin-orbit interaction (because of $S=0$). 
In fact, the spin-orbit interaction can mix the  good diquark, $S=0$, 
with the bad diquark, $S=1$, giving however only a second 
order correction to the mass that we provisionally neglect.
These considerations lead to:
\begin{equation}
M_Y=4160+B_c\left( \frac{L(L+1)}{2}\right)
\end{equation}
which leaves $\sim 100$~MeV for the orbital term, 
the only new ingredient with 
respect to Ref.~\cite{xparticles}. We try different ways to estimate $B_c$
from the corresponding terms in $q\bar{q}$ spectrum.
We find somewhat different results, which gives an idea of the
theoretical error involved.

We describe the masses of 
the $S=1,L=0,1$ states $\rho(770), a_1(1230), a_2(1320)$ with the equation:
\begin{equation}
M(S=1,L,J)=K + 2 A_q \vec{S}\cdot\vec{L}+ B_q \frac{L(L+1)}{2}.
\end{equation}
One finds at once:
\begin{equation}
B_q=\frac{a_1+a_2-2\rho}{2}=0.495~\rm{GeV}.
\end{equation}

For charm and beauty we take the difference between the lowest
$S=1,L=0$ state and the
center of $S=1,L=1$ mass spectrum and find:
\begin{equation}
B_{J/\psi}=425~{\rm MeV};\;\;\; B_{\Upsilon}=440~{\rm MeV}.
\end{equation}
For the quantum rotator $B\propto (mR^2)^{-1}$, 
with $R$ the radius of the bound 
state. Assuming the same radius and using $m_c=1.3$~GeV 
and $m_{[cs]}=2.1$~GeV as given above, we obtain from the light quark
case:
\begin{equation}
B_{c}=\frac{m_{q}}{m_{[cs]}}\times 495\simeq 120~{\rm MeV}
\end{equation}
(scaling from charmonium we would get $B_c\simeq 260$~MeV). 

An extreme alternative
is to consider the diquark as a single 
constituent quark and scale the orbital terms
as appropriate for Coulomb bound states. In this case, $B$ scales
like~\cite{martin} 
$(R^2 M)^{-1}$ and $R=(\alpha_s M)^{-1}$ so that 
\begin{equation}
B\propto\alpha_s^2 M. 
\end{equation}
This
formula does not reproduce the values of  $B_q,B_{J/\psi},B_{\Upsilon}$
simultaneously. Using $\Lambda_{\rm QCD}=190$~MeV we find $B_{J/\psi}\simeq
340$~MeV, $B_{\Upsilon}\simeq 500$~MeV; for 
a slightly larger $\Lambda_{\rm QCD}=270$~MeV we find  $B_{J/\psi}\simeq
135$~MeV, $B_{\Upsilon}\simeq 170$~MeV. In correspondence 
$B_{c}\simeq 370$~MeV and $134$~MeV respectively.
The experimental $Y$ mass clearly prefers a 
wider structure than charmonium but otherwise the orbital excitation 
picture is compatible within large theoretical errors: 
\begin{equation}
M_Y^{\rm th.}=4330\pm 70~{\rm MeV}.
\end{equation}

Given the quantum numbers $J^{PC}=1^{--}$, the state in Eq.~\eqref{eq:uno}
should decay strongly into a pair of mesons with open charm. The quark 
composition in~\eqref{eq:uno} implies a definite  preference for charm-strange
states:
\begin{equation}
\Gamma_Y(D_s\bar{D}_s)>>\Gamma_Y(D\bar{D})
\label{eq:tre}
\end{equation}
Dominant $D_s\bar{D}_s$ decay is quite a distinctive signature of the 
validity of the present model. 

Quark diagrams corresponding to 
the $D_s\bar{D}_s$ and to the $J/\psi\; f_0(980)$ decays are reported 
in Fig.~1. 
\begin{figure}[ht]
\begin{center}
\epsfig{
height=7truecm, width=7.5truecm,
        figure=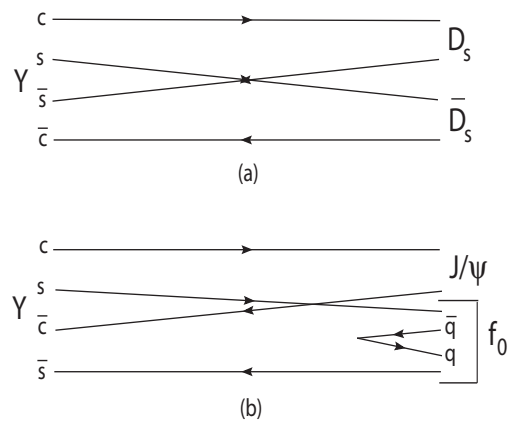}
\caption{\label{fig.1} \footnotesize 
(a) Quark diagram for the dominant decay channel to
$D_s\bar{D}_s$ see Ref.~\cite{prl}.
(b) Decay amplitude for $Y\to J/\psi f_0(980)$ 
under the assumption
that both $Y$ and $f_0$ are four-quark states.
}
\end{center}
\end{figure}
\vspace{-4truecm}
\begin{figure}[ht]
\begin{center}
\epsfig{
height=7truecm, width=7.5truecm,
        figure=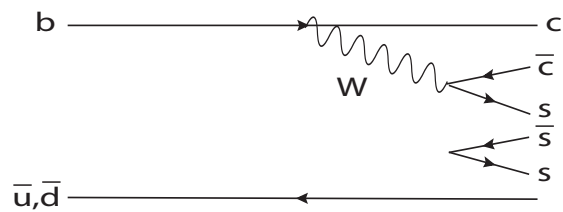}
\caption{\label{fig.1} \footnotesize 
Quark diagram for the weak decay of a $B^{-,0}$ meson 
into $YK^-$ and $YK_S$. Kaons can be obtained in two 
independent way by combining the spectator antiquark 
with strange quark from the weak vertex or from the sea pair.}
\end{center}
\end{figure}
Unlike the case of the $X(3872)$, the latter decay is not the 
dominant one.
Assuming a partial width similar to the total width of $X(3872)$, namely 
a few MeV's, 
one predicts a branching ratio for the $J/\psi f_0(980)$ channel 
in the order of $10^{-1}\div 10^{-2}$. The observation of BaBar, 
Eq.~\eqref{eq:babar}, therefore implies
for the $Y(4260)$ a leptonic width of $50\div 500$~eV, which is not 
unlikely for the one-photon 
production of such a complex state and consistent with the non-observation
of this resonance in multihadron $e^+e^-$ production around 
$E=4$~GeV~\cite{pdg}.

The $Y(4260)$ should be seen in $B^-$ and $B^0$ weak non-leptonic 
decays, see the quark diagrams in Fig.~2, with:
\begin{equation}
\Gamma(B^0\to YK_S)=\frac{1}{2}\Gamma(B^-\to YK^-).
\label{eq:quattro}
\end{equation}
Replacing the strange quark/antiquark with light quarks/antiquarks one obtains
a full nonet of $J^{PC}=1^{--}$ mesons. From the charm baryon spectrum
one finds~\cite{xparticles}: $(\kappa_{cs})_{\bf{\bar{3}}}\simeq
(\kappa_{cq})_{\bf{\bar{3}}}$, so that
the levels in the nonet are equispaced by $\simeq 185$~MeV (s=strangeness):
\begin{equation}
M_{Y(I=0,1;s=0)}=3.91~{\rm GeV};\;\;\;M_{Y(I=1/2;s=\pm 1)}=4.10~{\rm GeV}.
\end{equation}
The neutral members of the non-strange complex should be seen
in $e^+e^-$ annihilation and in $B$ non-leptonic decays(produced
by diagrams like
that in Fig.~2 with the $s\bar{s}$ pair replaced by $u\bar{u}$ or $d\bar{d}$). 
Dominant decay modes are in $D\bar{D}$.
Similar to the $X(3872)$ case, a significant isospin breaking in the 
wave function of the non-strange states can be expected. This should reflect
in unequal branching ratios of each mass eigenstate in $D^+D^-$ 
versus $D^0\bar{D}^0$.
In the limiting case of pure $\left([cu][\bar{c}\bar{u}]\right)_{\rm{P-wave}}$
and $\left([cd][\bar{c}\bar{d}]\right)_{\rm{P-wave}}$ the first would decay
in $D^0\bar{D}^0$ only and the second in $D^+D^-$. 
Decays into $J/\psi\pi^+\pi^-$ are 
expected to occur as well, with 
$\pi^+\pi^-$ peaking at the $\sigma(480)$ mass 
(restricted to the $I=0$ state, if isospin would be conserved).

The BaBar data suggest, although inconclusively, that there may be 
a considerably more narrow satellite line at a mass $M\sim 4330$~MeV. We
observe that this mass difference is of the order of the 
spin-spin interaction.
Indeed, if one calls into play bad diquark states with $S=1$ there are 
several additional $1^{--}$ states with the same quark composition, 
$(cs)(\bar{c}\bar{s})$. Among them, the state with both 
diquark and antidiquark 
spins in $S=1$, combined to $S_{\rm tot}=2$.
This state projects only on spin one $c\bar{s}$ and 
$s\bar{c}$ states. 
In the (not unrealistic) limit where the spin of the $s$ quark 
is a good quantum number, such state could decay only into $D_s^*\bar{D}_s^*$
pairs, with substantial reduction of its decay width.

{\sl Acknowledgments}. 
We wish to thank R. Faccini and A. Martin for useful information 
and interesting discussions. This work was
partially supported by MIUR (Italian Minister of Instruction University and
Research) and INFN (Italian National Institute for Nuclear Physics).


\begin{thebibliography}{99}
\bibitem{xparticles} L.~Maiani, F.~Piccinini, A.D.~Polosa, V.~Riquer,
Phys.~Rev.~{\bf D71}, 014028 (2005). 
\bibitem{babar} BaBar Collaboration, [arXiv:hep-ex/0506081].
\bibitem{prl}L.~Maiani, F.~Piccinini, A.D.~Polosa, V.~Riquer,
Phys. Rev. Lett. {\bf 93}, 212002 (2004). 
\bibitem{jaffe} See e.g. R.L.~Jaffe, Phys. Rept. {\bf 409}, 1 (2005).
\bibitem{martin} R.A.~Bertlmann, A.~Martin, 
Nucl. Phys. {\bf B168}, 111 (1980).
\bibitem{pdg} Particle Data Group, S.~Eidelman {\it et al.}, Phys. Lett. 
{\bf B592}, 1 (2004).
\end{thebibliography}
\end{document}